\documentstyle[aps,prb,epsfig,twocolumn,floats,psfig]{revtex}

\begin{document}
\draft
\title{Theory for Phase Transitions in Insulating V$_2$O$_3$}
\author{A. Joshi,$^{1}$ Michael Ma,$^{1}$ F. C. Zhang$^{1,2}$}
\address{$^{1}$Department of Physics, University of Cincinnati, OH 45221-0011}
\address{$^{2}$Department of Physics, Chinese University of Hong Kong, 
Shatin, Hong Kong} 
\date{\today}
\maketitle

\begin{abstract}
We show that the recently proposed $S=2$ bond model with orbital degrees
of freedom for insulating V$_{2}$O$_{3}$ not only explains the anomalous
magnetic ordering, but also other mysteries of the magnetic phase
transition. The model contains an additional orbital degree of freedom that
exhibits a zero temperature quantum phase transition in the Ising
universality class.
\end{abstract}



\smallskip

\narrowtext

The metal-insulator transition in vanadium oxide (V$_{2}$O$_{3}$) has long
been hailed as a physical realization of the Mott transition~\cite
{mcWhan1,mcWhan2,mott}. However, the insulating phase displays essential
deviations from the standard Mott insulator with only spin degrees of
freedom. For example, while it undergoes a transition from paramagnetic to
antiferromagnetic (AF) insulator with decreasing temperature, the magnetic
ordering pattern observed~\cite{moon,word,bao1} (hitherto called RS) is not
that of the simplest two-sublattice Neel state (hitherto referred to as AS).
Some time ago,
Castellani, Natoli, and Ranninger (CNR)~\cite{castellani} proposed a model in
which the low energy degrees of freedom for each V ion are a spin ($S=1/2$)
coupled to doubly degenerate orbital. They showed that within certain
parameter regime, the ground state exhibits the RS magnetic ordering
pattern together with orbital ordering. Recently, resonant x-ray scattering
experiments~\cite{paolasini} have observed orbital ordering in V$_{2}$O$_{3}$
with an ordering wavevector consistent with one of the ordered state (so-called
RO state) obtained by CNR.

The CNR model relies on one out of the two electrons on a V ion forming a
spin singlet bond with its counterpart on the neighboring V ion along the
c-axis, leaving one electron occupying doubly degenerate orbitals on each V
ion. This has recently been criticized for not respecting the strong on-site
coulomb repulsion and Hund's rule coupling~\cite{park,ezhov}. This
criticism is supported by polarized soft x-ray experiments~\cite{park}, and strongly
suggests that the effective spin on a V ion should have $S=1$ instead of 
$S=1/2$. Recently, a $S=1$ model incorporating the orbital degree of freedom
has been introduced~\cite{mila}. In this model V-V pair along c-axis is
locked ferromagnetically into a total spin 2 ``bond'' with a double orbital
degeneracy. These bonds occupy a lattice that is topologically equivalent to
a cubic lattice (Fig. 1). Virtual in-plane hopping in the honeycomb planes
couple such bonds to their neighbors. For Hund's rule coupling and hopping
constants consistent with reported/expected values, it was shown that the
ground state has RS magnetic and ferro-orbital (FO) order. The FO ordering
was shown to be also consistent with the resonant x-ray experiments.

\begin{figure}[h]
\centering\epsfig{file=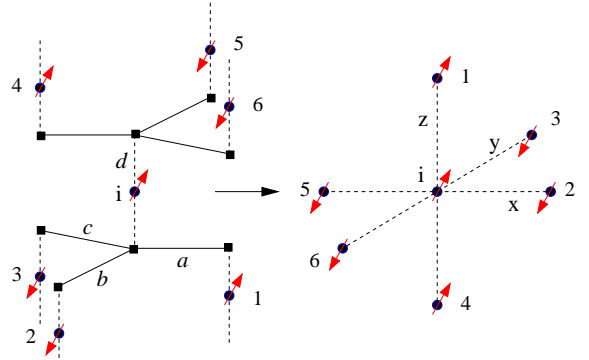,width=7.5cm,angle=0}
\caption{Mapping of bonds on the corundum lattice to sites on a
cubic lattice~\cite{mila,shiina}. The positions of V ions are shown 
by squares and the centers
of the bonds by circles. The arrows represent the spin of 
V-ions in V-V pair in the RS state.}
\label{cubic}
\end{figure}

In addition to the anomalous RS magnetic order, there are other mysteries
associated with the V$_{2}$O$_{3}$ insulator~\cite{rice}: I) the magnetic
and orbital ordering occur at the same transition temperature~\cite
{paolasini}, in contrast to the Jahn-Teller transitions observed in other
transition metal oxides like MnO; II) the transition is strongly first 
order~\cite{bao2}; and III) above the transition
temperature, neutron scattering observes a relatively broad peak in the
susceptibility for wavevector of AS ordering but no features for the RS
ordering wavevector~\cite{bao2}.

In this letter, we will show that the $S=2$ bond model~\cite{mila} gives
satisfactory explanation for the three mysteries above. In addition, we will
elucidate on a new orbital degree of freedom, which can lead to a second orbital 
transition at a lower
temperature. The critical temperature of this second orbital transition goes
to zero as the intraplane hopping approaches some critical value, giving
rise to a zero temperature quantum phase transition. This second transition is
well described by the transverse field Ising model.

We begin by briefly recalling the steps leading to the effective $S=2$ bond
Hamiltonian. The large on-site, same-orbital coulomb repulsion ($U\approx 5eV$)
and Hund's coupling ($J\approx 0.6-1eV$)~\cite{park,ezhov} constrains the 
Hilbert space of
each V ion to that of 2 electrons with total spin $S=1$, hence singly
occupying two of the lower three $t_{2g}$ d-orbitals favored by crystal
field. The remaining 9-fold degeneracy of a single ion is then
lifted due to virtual hopping. Let $t_{ij}^{\delta }$ be the various hopping
integrals between nearest neighbor (nn) V ions, where $\delta =a,b,c,d$ stands
for the direction of the bond (Fig. 1) and $(i,j)=1,2,3$ for the orbitals 
$e_{g1}$, 
$e_{g2}$, $a_{1g}$ respectively. Since $t_{33}^{d}$ is by far the largest,
the strongest effect is due to virtual hopping between vertical pairs of V
ions. When combined with $J$ large, this further reduces the Hilbert space
of a single pair to that of total spin $S=2,$ with an occupation of ($e_{g1}$,
 $e_{g2})$ by the two electrons on one V ion and ($e_{g}$, $a_{1g})$
on the other. The orbital configuration proposed here is qualitatively
consistent with the polarized x-ray absorption spectra~of V$_{2}$O$_{3}$\cite
{park}. The basis for such a V-ion pair can then be written as $\left|
\sigma _{z}\tau \mu \right\rangle $. Here $\sigma _{z}=-2,-1,...2$ is the
spin state, $\tau =+1$ $(-1)$ if $e_{g1}$ ($e_{g2}$ ) is occupied on the ion
with ($e_{g}$, $a_{1g})$ occupation, and $\mu =+1$ $(-1)$ denotes whether
the top V ion has the ($e_{g1}$, $e_{g2})$ or ($e_{g}$, $a_{1g})$
occupation.

Within this restricted Hilbert space, we can now obtain the effective
Hamiltonian by considering the lifting of the degeneracies due to the
effects of the other $t_{ij}^{\delta }$'s. The effective Hamiltonian is
expressed by defining on each vertical pair a spin 2 operator ${\bf \sigma }
, $ a psuedospin 1/2 operator ${\bf \tau },$ and yet another psuedospin
operator ${\bf \mu },$ which act on the $\sigma _{z},\tau, $ and $\mu $
degrees of freedom respectively. The effective Hamiltonian must respect
global $SU(2)$ symmetry in spin space, $Z_{3}$ symmetry in $\tau $ space,
and $Z_{2}$ symmetry in $\mu $ space. The $Z_{3}$ Potts symmetry in $\tau $ 
reflects
the $C_{3}$ rotational symmetry of the corundum lattice about the c-axis
while the Ising $Z_{2}$ symmetry in $\mu $ reflects the global
``top-bottom'' inversion symmetry of the V-ion pairs. We consider only
processes where after two (virtual) hops, the system is restored into the
restricted Hilbert space. With no loss of generality, we assume for $
t_{ij}^{a}$ that only $t_{11}$ and $t_{23}$ are non-negligible.

At temperature $\approx$ 155K, x-ray experiment has shown that all V ions
remain equivalent in the AF phase~\cite{dernier}, implying no symmetry
breaking of the $Z_{2}$ $\mu $ symmetry at the magnetic transition. If we
further neglect fluctuation effects, effectively setting $\mu _{z}=0$ in the
effective Hamiltonian, $\mu $ becomes decoupled from the ${\bf \sigma }$ and
${\bf \tau }$ degrees of freedom, and the Hamiltonian is simplified to \cite
{mila,shiina}

\begin{eqnarray*}
H=a_{0}\sum_{\left\langle ij\right\rangle }{\bf S}_{i}\cdot {\bf S}%
_{j}+a_{1}\ \sum_{\left\langle ij\right\rangle }{\bf S}_{i}\cdot {\bf S}_{j}(
{\bf \tau }_{i}\cdot \widehat{{\bf n}}_{ij}+{\bf \tau }_{j}\cdot \widehat{%
{\bf n}}_{ij})   \\
+ a_{2}\sum_{\left\langle ij\right\rangle }{\bf \tau }_{i}\cdot \widehat{%
{\bf n}}_{ij}{\bf \tau }_{j}\cdot \widehat{{\bf n}}_{ij}+b_{2}\sum_{\left%
\langle ij\right\rangle }{\bf S}_{i}\cdot {\bf S}_{j}\left( {\bf \tau }%
_{i}\cdot \widehat{{\bf n}}_{ij}{\bf \tau }_{j}\cdot \widehat{{\bf n}}%
_{ij}\right)   \\
\end{eqnarray*}
\noindent where $\widehat{{\bf n}}_{ij}=\widehat{{\bf n}}_{1},$ $\widehat{%
{\bf n}}_{2}, $ $\widehat{{\bf n}}_{3}$ for $i,j$ nn in the $x,
$ $y,$ $z $ directions respectively. The $\widehat{{\bf n}}_{i}$'s are unit
vectors in the $x-z$ plane of the $\tau$ space, with $\widehat{{\bf n}}_{3}=\widehat{z},$ while $%
\widehat{{\bf n}}_{1}$ and $\widehat{{\bf n}}_{2}$ are rotated from $%
\widehat{{\bf n}}_{3}$ by 120$^{0}$ and 240$^{0}.$ The coupling constants $%
a_{0}$ etc. depend on $J$ and $\gamma =\left( t_{11}/t_{23}\right) ^{2}$.
The model contains three types of local order parameters: a vector order
parameter ${\bf M}_{i}=<{\bf S}_{i}>$,
a Pott's type order parameter $r_{i}^{\alpha }=<\tau _{i}\cdot \widehat{{\bf %
n}}_{\alpha }>$, and a tensor order
parameter ${\bf Q}_{i}^{\alpha }=<{\bf S}_{i}\tau _{i}\cdot \widehat{{\bf n}}%
_{\alpha }$ $>$. Assuming up to four sublattice
ordering in these three order parameters on the equivalent cubic lattice
with wavevectors $k_{x}=k_{y}=0,\pi ,$ $k_{z}=0,\pi $, we approximate $H$ by
a single-site Hamiltonian using usual mean field theory (MFT) to find the values
of ${\bf M}_{i},$ $r_{i}^{\alpha },$ and ${\bf Q}_{i}^{\alpha }$ that
minimizes the free energy. The phase diagram for zero temperature thus
obtained is shown in Fig. 2. 

\begin{figure}[h]
\centering\epsfig{file=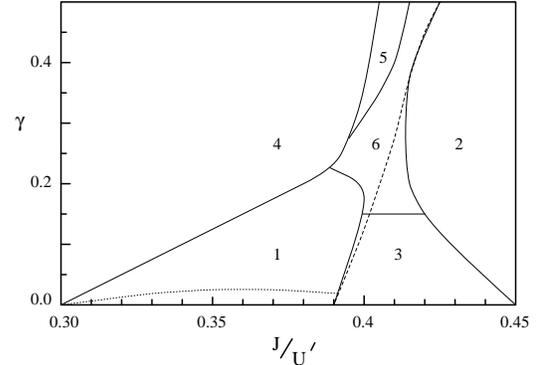,width=7.5cm,angle=0}
\caption{$T=0$ phase diagram of the $S=2$ bond model. $U' = U - 2J$. Phases 
are 1) FORS, 2) AOFS,
3) FORS$^{\prime }$, 4) AOAS, 5) AO$_{z}$RS, and 6) AO$_{x}$RS$^{\prime }$
(subscripts after AO indicate direction of AO ordering in $\tau $ space).
Dashed line shows where $a_{0}=0.$  Dotted line 
shows the boundary between
first order (below the line) and second order transitions for the FORS phase.}
\label{phasediagram}
\end{figure}

The phase diagram can be understood
qualitatively as follows\cite{shiina,shiina1}. The key dependences of the
various coupling constants on $J$ and $\gamma $ are that $a_{0}$ changes
from $>0$ to $<0$ as $J$ is increased, $a_{1}<0$ everywhere, and $
a_{2},b_{2}=0$ for $\gamma =0$ and $>0$ otherwise, with $a_{2}\geq 4b_{2}.$
The $a_{0}$ term favors AS or ferromagnetic spin (FS) ordering while the
$a_{2}$ and $b_{2}\ $ terms combined favor antiferro-orbital (AO) ordering.
The $a_{1}$ term in $H$ appears as an explicit symmetry breaking field on
$\tau .$ However, for the AS or FS states or any other state whose spin
correlations $<{\bf S}_{i}\cdot {\bf S} _{j}>$ do not break the lattice C3
symmetry, this field equals zero when summed over all bonds, and the $a_{1}$
term gives no contribution to the energy. Thus, close to the boundary where
$a_{0}$ changes sign, the spins should order magnetically in a way to break
the C3 symmetry to take advantage of the $a_{1}$ term. This accounts for the
states RS and RS$ ^{^{\prime }}$ which differ from each other by having 1
ferromagnetic bond and 2 antiferromagnetic bonds on the honeycomb plane for
the RS and vice versa for RS$^{^{\prime }}$. Choosing the ferromagnetic bond
for RS and antiferromagnetic bond for RS$^{^{\prime }}$ to be along the $a$
direction of the honeycomb planes, these magnetic orderings set up an uniform
field on $\tau $ in the $+z$ and $-z$ directions respectively, thus for small
$\gamma $ causing $\left\langle \tau _{iz}\right\rangle $ to become uniformly
non-zero (FO phase).

Of these phases, the one that is consistent with the magnetic and orbital
ordering observed by neutron scattering and resonant x-ray scattering
experiments is the FORS phase. Below we shall focus on this phase and study
the finite temperature phase transition of the model to explain the three
mysteries of insulating V$_{2}$O$_{3}$.

I) In V$_{2}$O$_{3},$ the magnetic transition is accompanied by orbital
ordering. In our MFT, we indeed find the spin and orbital to order at the
same temperature. This result is robust, and can be explained more generally
as a consequence of the form of the Hamiltonian together with FORS type
ordering. Since
the RS ordering produces a symmetry breaking field on $\tau ,$ any RS
ordering must necessarily be accompanied by non-zero $\left\langle \tau
_{i}\right\rangle $'s. The converse of this is not true, and
orbitals can in principle order without spin ordering.\ However, the only term
in $H$ that can cause FO ordering is the $a_{1}$ term which vanishes without RS
ordering\cite{pierels}. Thus, we conclude that for FORS (or FORS$^{^{\prime
}}$) ordering, magnetic and orbital ordering must occur together. This is
not true for the other ground states. For AOAS or AOFS, as $T$ is lowered,
magnetic ordering can first occur without orbital ordering or vice versa.
For AORS and AORS$^{^{\prime }}$, magnetic ordering must be accompanied or
preceded by orbital ordering, but orbital ordering can occur without
magnetic ordering.

We should note that the observed monoclinic distortion below $T_{c}$ plays
no role in our theory. Indeed, we believe it to be a by-product of the FORS
transition breaking the $C_{3}$ rotation symmetry of the honeycomb planes.

II) The magnetic transition is strongly first order, with an entropy jump of
$\approx k_{B}\ln 2$ per V ion. The size of entropy jump suggests that the
first order transition should not be induced by critical fluctuations, but
should be mean field in origin. Indeed, our MFT
shows a first order transition for the paramagnetic-FORS transition for small
$\gamma$ (Fig. 2).
It should be remarked that while the symmetry of $H$ allows
third order invariants in the Landau-Ginzburg free energy functional of the
form $r^{3}$ and $MQr,$ they are not the cause of the first order transition.
This is because the FO susceptibility at $T_{c}$ is small since the FO ordering
is not favored by the bare orbital-orbital coupling of $H$.
Indeed, for values of  $\gamma $ such that the transition
into FORS phase is continuous, $r\sim MQ$ below but close to the transition.
Instead, the first order transition is due to the presence of a negative
$M^{2}Q^{2}$ quartic term.

While MFT gives a first order transition, the calculated entropy jump and 
maximum value of $\gamma$ are rather small, especially when compared to the 
actual experimental value of $\approx k_{B}\ln 2$ per V ion. We believe this 
to be a result of ignoring
nn AF correlation in the paramagnetic state in a MFT for the
FORS transition. This correlation is seen to be sizeable experimentally (see
III below), and serves to stabilize the paramagnetic state. We estimate the 
nn correlation using high
temperature expansion to second order in $a_{0}$ and find the entropy jump
and maximum $\gamma$ to be significantly increased without much reduction in 
$T_{c}$. A schematic
of the free energy functional without and with nn AF correlations included
is shown in Fig. 3.

\begin{figure}[h]
\centering\epsfig{file=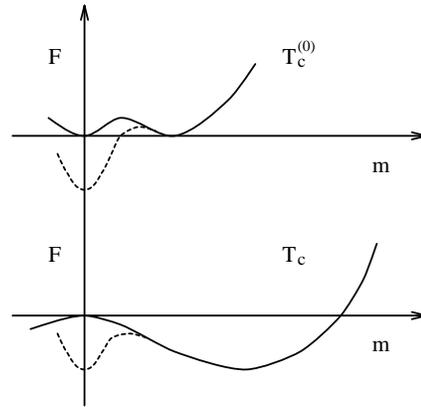,width=5.5cm,angle=0}
\caption{Solid line shows schematically the MF free energy as a function of the
FORS order parameter. Dotted line shows the free energy when nearest
neighbor AF correlations are included for the paramagnetic state. Inclusion
of such correlations stabilizes the PI state 
and lowers the transition temperature slightly from $T_{c}^{(0)}$ to $T_{c}$
while significantly increasing the order parameter and entropy jump.}
\label{fenergy}
\end{figure}

III) Given the magnetic RS ordering below $T_{c},$ one might expect peaks at
the three symmetry related RS wavevectors above $T_{c},$ and that these
peaks would sharpen as the temperature is lowered towards $T_{c}.$ Neutron
scattering~\cite{bao2} above $T_{c},$ however, shows a broad peak in AS
wavevector but no feature in the RS wavevectors. Moreover, as $T$ decreases,
the broad peak increases in intensity but does not sharpen appreciably,
indicating that nn spins are increasingly AF correlated but the correlation
length remains of order of nn. This seemingly surprising feature can be
explained qualitatively as follows in our theory. Above $T_{c}$, there are
fluctuations in both the AS and RS states. The nn spin correlation of AS is
of course AF. While the correlation of RS for a particular symmetry broken
state has a pattern with ferromagnetic in one nn bond and AF in other two nn
bonds in the honeycomb plane, the nn correlation when averaged over the
three RS states also is AF. For long distance correlations, however, the
signs of correlations due to fluctuations of AS and RS states can be
competing. As a result, these spin correlations remain weak.

FO ordering exists only for RS or RS$^{^{\prime }}$ ordering. Also,
the simultaneity of magnetic and orbital ordering, and the first order
phase transition depend crucially on FORS ordering. Thus, the phase transition
phenomenology can be changed by changing the type of magnetic ordering.\ This
can be achieved by applying an uniform magnetic field which with increasing
strength will change the preferred magnetic ordering from RS to FS. This will
in turn change the orbital ordering and the phase transition properties. This
can be observed by performing resonant X-ray scattering in the presence of a
magnetic field.

Let us now return to the $\mu $ degree of freedom. At temperature
significantly lower than $T_{c}$ , the FORS ordering will be approximately
saturated and frozen out, and $\mu $ becomes the only relevant degree of
freedom. The effective Hamiltonian thus becomes
\[
H_{\mu }=-\Delta \sum_{i}\mu _{ix}+\sum_{\left\langle ij\right\rangle
}J_{ij}\mu _{iz}\mu _{jz}
\]
where $\Delta =2t_{11}^{d}t_{33}^{d}/(U-3J),J_{ij}=J_{1}$ for
ferromagnetically aligned nn, and $J_{ij}=J_{2}$ for
antiferromagnetic ones. $J_{1}$ and $J_{2}$ are given by
\begin{eqnarray*}
J_{1} &=&\frac{t_{23}^{2}}{U-3J}, \\
J_{2} &=&\left( \frac{2}{(U-3J)}+\frac{1}{U}-\frac{3}{U+2J}\right) \frac{%
t_{23}^{2}}{24}(1-\frac{9}{8}\gamma )
\end{eqnarray*}
We recognize this Hamiltonian as the transverse field Ising model (TFIM). At
$T=0,$ this model has a quantum phase transition from a disordered state for
large $\Delta $ to an ordered one for small $\Delta .$ The ordering pattern
depends on the signs of $J_{1}$ and $J_{2}$.
We see that $J_{1}>0$ always, while $J_{2}$ changes sign as $%
\gamma $ is decreased. For $J_{2}<0,$ $\mu $ ordering does not lead to
further reduction of translational invariance from the FORS state, and no
new wavevector $q$ will be observed in resonant x-ray diffraction, only
intensity change. For $J_{2}>0,$ there will be further change in Bravais
lattice symmetry, and peaks at new $q$'s will be observed. The value of $%
\gamma $ necessary for FORS ordering and in particular to have a first order
transition for estimated value of $J$ suggests that $J_{2}>0.$

For the 3D TFIM, its finite temperature transition is in the same
universality class as the 3D Ising model and its T=0 transition in that of
the 4D Ising model respectively. However, if $J_{2}$ is close to zero, then
the system behaves like a quasi-$1D$ one. Therefore, as $J_{2}$ varies, it
may be possible to observe a dimensional crossover. If $J_{2}$ is not too
small, the critical value of $\Delta $ for LRO may be obtained by
MFT as $2J_{1}+4\left| J_{2}\right| =\Delta /\tanh (\Delta /k_{B}T)$. This
implies the critical temperature $T_{\mu}$ rises sharply from $0$ as $\Delta$
is reduced from its $T=0$ critical value. The MF value of $T_{\mu }$ should
be accurate provided $\Delta $ is neither too close to $\Delta _{c}$ nor too
small. For $\Delta $ close to $\Delta _{c}$, $T_{\mu }$ is governed by
quantum critical phenomena, and scales as the inverse correlation time, so
that up to logarithmic corrections $T_{\mu }\sim \xi _{t}^{-1}\sim \left(
\Delta _{c}-\Delta \right) $. For small enough $\Delta $, $T_{\mu }$ can
become comparable to $T_{c},$ and the assumption of saturated FORS ordering
will not be valid.

We would like to acknowledge H. R. Krishnamurthy, T. M. Rice, R. Shiina, 
F. Mila, Rajiv Singh, Wei Bao, T. Ziman and X. Hu for useful discussions. 
M. Ma acknowledges the hospitality of Hong Kong University of Science and
Technology. This
work  was supported in part by DOE Grant No. DE/FG03-98ER45687 and a
URC student fellowship of the University of Cincinnati (A. Joshi).


\begin{references}
\bibitem{shiina1}  Details of the $T=0$ phase diagram using a similar
calculation has been previously reported\cite{shiina}.
\bibitem{pierels}  Beyond mean field, there is also the possibility of
orbital ordering along with RS spin correlations but not ordering, in effect
an orbital driven spin Peirels transition.
\end{references}
\end{document}